\documentclass[preprint,aps,floats,amssymb,showpacs]{revtex4}
\usepackage{epsfig}

\newcommand{\be}{\begin{equation}}
\newcommand{\ee}{\end{equation}}
\newcommand{\bea}{\begin{eqnarray}}
\newcommand{\eea}{\end{eqnarray}}
\newcommand{\bean}{\begin{eqnarray*}}

\newcommand{\eean}{\end{eqnarray*}}
\font\upright=cmu10 scaled\magstep1 \font\sans=cmss10
\newcommand{\ssf}{\sans}
\newcommand{\stroke}{\vrule height8pt width0.4pt depth-0.1pt}
\newcommand{\Z}{\hbox{\upright\rlap{\ssf Z}\kern 2.7pt {\ssf Z}}}

\newcommand{\C}{{\rlap{\rlap{C}\kern 3.8pt\stroke}\phantom{C}}}
\newcommand{\R}{\hbox{\upright\rlap{I}\kern 1.7pt R}}
\newcommand{\CP}{\C{\upright\rlap{I}\kern 1.5pt P}}
\newcommand{\PP}{\hbox{\upright\rlap{I}\kern 1.5pt P}}

\newcommand{\identity}{{\upright\rlap{1}\kern 2.0pt 1}}

\newcommand{\HH}{\mbox{\hbox{\upright\rlap{I}\kern 1.7pt H}}}

\newcommand{\zb}{{\bar z}}

\newcommand{\fr}{\frac}

\newcommand{\ra}{\rightarrow}
\newcommand{\al}{\alpha}
\newcommand{\bt}{\beta}
\newcommand{\pr}{\partial}
\newcommand{\hs}{\hspace{5mm}}

\newcommand{\dg}{\dagger}

\newcommand{\acc}{\\[3mm]}

\begin{document}




\title{Harmonic
map analysis of $SU(N)$ gravitating Skyrmions}
\renewcommand{\thefootnote}{\fnsymbol{footnote}}
\author{Yves Brihaye \footnote{yves.brihaye@umh.ac.be}}
\address{Facult\'e des Sciences, Universit\'e de Mons, 7000 Mons, Belgium}
\author{Betti Hartmann \footnote{b.hartmann@iu-bremen.de}}
\address{School of Engineering and Sciences, International
University Bremen (IUB), 28275 Bremen, Germany}
\author{Theodora Ioannidou \footnote{T.Ioannidou@ukc.ac.uk}}
\address{Institute of Mathematics,  University
of Kent, Canterbury CT2 7NF, UK   }
\author{Wojtek Zakrzewski\footnote{W.J.Zakrzewski@durham.ac.uk}}
\address{Department of Mathematical Sciences, University
of Durham, Durham DH1 3LE, U.K.}
\date{\today}
\setlength{\footnotesep}{0.5\footnotesep}
\begin{abstract}
In this paper the $SU(N)$ Einstein-Skyrme system is considered. 
We express  
the chiral field (which is not a simple embedding of the $SU(2)$ one) in terms 
of harmonic maps.
In this way, $SU(N)$ spherical symmetric equations can
be obtained easily for any $N$ and the 
gravitating skyrmion solutions of these equations can be 
studied.
In particular, the  $SU(3)$ case is considered in detail and  three
different types of gravitating skyrmions with topological
charge $4$, $2$ and $0$, respectively, are constructed numerically.
Note that the configurations with topological charge $0$ correspond
to mixtures of skyrmions
and antiskyrmions. 
\end{abstract}
\maketitle
\renewcommand{\thefootnote}{\arabic{footnote}}

\section{Introduction}
Nonlinear field theories coupled to gravity have received a lot
of interest in the past decade. It has been discovered that 
gravitational interaction may lead  to genuinely nonperturbative
phenomena like gravitationally bound configurations
of nonabelian fields, while
the study of black hole solutions in various models revealed the 
possibility of nonlinear hair on black holes \cite{LW} which
questioned the validity of the unqualified no-hair conjecture.

One of the candidates of such investigations is the Einstein-Skyrme model which can e.g. describe
the interaction between a baryon and a black hole 
(a configuration which might have been produced in the very early universe).
So far, most of the studies have concentrated on the $SU(2)$ Einstein-Skyrme
model \cite{LM,DHS,BC}.
 In particular, in  \cite{LM} it has been  shown that the Schwarzschild
 black hole  can support chiral (``Skyrme") hair and it has been argued that 
such configurations 
might be stable.
The presence of the horizon in the core of the skyrmion unwinds the skyrmion, 
leaving fractional baryon charge outside the horizon. 
More systematic investigations of the model were undertaken in \cite{DHS} 
by solving numerically the static spherically symmetric Einstein-Skyrme 
equations. Globally regular solutions with baryon number one \cite{BC} and
black holes with chiral hair were found and their stability properties were
studied in detail \cite{DHS,BC}.

The first examples of  nonembedded solutions for a higher group, namely the $SU(3)$ group,
were the $SO(3)$ solitons with even topological charge. The lowest energy
solution  corresponds to a bound state of two
gravitating skyrmions \cite{KKSO}. Specifically, it was found that the two branches
of these regular solutions exist, which merge at a critical
value of the gravitational coupling.

In this paper we consider particle-like solutions of the $SU(N)$ Einstein-
Skyrme model (for $N\geq 2$). In particular, we study the deformation of 
the multiskyrmion configurations \cite{IPZ2} (derived using the
 harmonic map  ansatz) when gravity is introduced. Note that the use
of harmonic maps for gravitating skyrmions can be traced  to
\cite{conor}. 
New types of solutions are found, which
 correspond to skyrmion-antiskyrmion configurations and have
topological charge $0$. Like the non-gravitating skyrmion-antiskyrmions,
these configurations are also saddle points of the energy
functional and thus are likely to be unstable.

Our paper is organised as follows: in Section II, we present the $SU(N)$ 
Einstein-Skyrme
model, while  in Section III we give the harmonic map ansatz.
In Section IV,
we present the spherically symmetric equations of motion and in Section V 
we discuss
our numerical results for the $SU(3)$ case. In this latter section,
we also point out that we recover the already known solutions for the topological
charge $4$ and $2$ \cite{KKSO}, but also derive new solutions describing
the gravitating skyrmion-antiskyrmions
with the topological charge $0$. Our conclusions
are summarised in Section VI.

\section{The $SU(N)$ Einstein-Skyrme Model}

The $SU(N)$ Einstein-Skyrme action reads:
 \be
  S=\int \left[\fr{R}{16\pi G}-\fr{1}{2} \,\mbox{tr}\left(K_\mu
  \,K^\mu\right)-\fr{1}{16}\,\mbox{tr}\left(\left[K_\mu,K_\nu\right]
\left[K^\mu,K^\nu\right]\right)\right]\sqrt{-g}\, d^4x
  \label{ac}
  \ee
where $K_\mu=\pr_\mu U U^{-1}$ for $\mu=0,1,2,3$ and $U(x^\mu)\in SU(N)$ 
is the matter field, $g$ denotes the determinant of the
  metric and
$G$ represents Newton's constant.
In order for the finite-energy configurations to exist the Skyrme field
has to go to a constant matrix at spatial infinity:  $U\ra I$ as
$|x^\mu|\ra \infty$.

To derive the classical equations of motion of our system we perform
the variation of the action (\ref{ac}) with respect to the metric and 
the Skyrme field.
The variation with respect to the metric
$g^{\mu \nu}$ gives the Einstein equations:
 \be
 R_{\mu \nu}-\fr{1}{2}g_{\mu \nu} R=8\pi G \,T_{\mu \nu},
 \label{En}
 \ee
 where $R_{\mu\nu}$ denotes the Ricci tensor and the stress-energy tensor
 $T_{\mu\nu}=g_{\mu\nu}{\cal L}-2\fr{\pr{\cal L}}
{\pr g^{\mu\nu}}$ is given by:
\be
T_{\mu\nu}\!\!=\mbox{tr}\!\left(K_\mu K_\nu-\fr{1}{2}g_{\mu \nu}
K_\al K^\al \right)+ \fr{1}{4} 
\mbox{tr}\!\left(g^{\al \bt}\left[K_\mu,K_\al\right]\left[K_\nu,K_\bt\right]-
\fr{1}{4}g_{\mu \nu} \left[K_\al,K_\bt\right]\left[K^\al,K^\bt\right]\right).
\label{T}
\ee
The variation of the action with respect to the matter fields
leads to the Euler-Lagrange equations  which we will discuss
 in the next section.

The Einstein-Skyrme system has a topological current which is covariantly 
conserved, yielding the topological charge \cite{GKK}:
\be
B=\int \sqrt{-g} \,B^0\, d^3x \ \ ,
\label{b}
\ee
where 
\be
B^\mu=-\fr{1}{24\pi^2\sqrt{-g}}\,\varepsilon^{\mu\nu\al\bt}\,\mbox{tr}\left
(K_\nu K_\al K_\bt\right)
\ee
and  $\varepsilon^{\mu\nu\al\bt}$ is the (constant) fully antisymmetric
tensor.

 In what follows we will concentrate our attention on studying the static
Einstein-Skyrme equations and we are going to construct
their static spherically symmetric solutions by using harmonic maps
 for the Skyrme field.

\section{Harmonic Map Ansatz}
The starting point of our investigation is the introduction of the
coordinates $r,z,\zb$ on $\R^3$. In terms of the usual spherical
coordinates $r,\theta,\phi$ the Riemann sphere variable $z$ is given by:
$z=e^{i\phi} \tan(\theta/2)$ and $\bar{z}$ is the complex conjugate
of $z$. In this system of coordinates the
Schwarzschild-like (spherically symmetric) metric reads:
\begin{equation}
ds^2=-A^2(r)C(r)\,dt^2+\fr{1}{C(r)}\,dr^2+
\frac{4r^2}{(1+|z|^2)^2}\,dz d\bar{z} ,\hs C(r)=1-\fr{2m(r)}{r},
 \label{s}
\end{equation}
where $m(r)$ is the mass function.
For this metric  the square-root of the determinant takes the simple form:
\begin{equation}
\sqrt{-g}=iA(r) \,\frac{2r^2}{(1+|z|^2)^2} \ .
\end{equation}
 After substituting
 the metric (\ref{s}), the action (\ref{ac}) becomes:
 \bea
 S&=&\int 
\bigg[\fr{R}{16\pi G}+ \mbox{tr}\bigg(
-\fr{1}{2}CK_r^2-\fr{(1+|z|^2)^2}
{2r^2}|K_z|^2+\fr{1}{32}\fr{(1+|z|^2)^4}{r^4}
\left[K_z,K_{\bar{z}}\right]^2\nonumber\\
&&\hs \hs\hs\hs\hs\hs -\fr{1}{8} \fr{(1+|z|^2)^2}{r^2}
C\bigg|\left[K_r,K_z\right]\bigg|^2\bigg)\bigg]
\sqrt{-g}\,dt\, dr \,dzd\bar{z}
  \label{ac1}
  \eea
while the baryon number is equal to
\be
B=-\fr{1}{8\pi^2}\int\mbox{tr}\left(K_r\left[K_z,K_{\bar{z}}\right
]\right) dr\, dz\,d\bar{z} \ .
\label{B}
\ee

 In addition, the Einstein equations (\ref{En}) take the form:
 \bea
\fr{2}{r^2}m'&=&8\pi G\,T^0_0\nonumber\\
\fr{2}{r}\,\fr{A'}{A}C&=&
8\pi G\,\left(T^0_0-T^r_r\right)
\label{ff}
 \eea
where the prime denotes the derivative with respect to $r$ and
\bea
 T^0_0&=&-\fr{C}{2}\,\mbox{tr}\left(K_r^2\right)
-\fr{ (1+|z|^2)^2}{2r^2}\,\mbox{tr}\left(|K_z|^2\right)
-\fr{1}{8} \fr{C(1+|z|^2)}{r^2}\,\mbox{tr}\left(\bigg|
[K_r,K_z]\bigg|^2
\right)
\nonumber\\
&&-\fr{1}{32}
\fr{(1+|z|^2)^4}{r^4}\,\mbox{tr}\left(\left[K_z,K_{\bar{z}}\right]^2
\right),\acc
T^0_0-T^r_r&=&-C\,\mbox{tr}\left(K_r^2\right)
-\fr{1}{4}\fr{C(1+|z|^2)^2}
{r^2}\mbox{tr}\left(\bigg|[K_r,K_z]\bigg|^2\right).
\eea

 Following \cite{IPZ2} the application of the harmonic map ansatz to
describe the matter fields, 
corresponds to setting:
 \bea
U&=&\exp\left\{2i\sum_{i=0}^{N-2}g_i\left(P_i-\fr{I}{N}\right)\right\}
\nonumber\acc
&=&e^{-2ig_0/N}\left(1+A_0\,P_0\right)e^{-2ig_1/N}\left(1+A_1\,P_1\right)
\dots e^{-2ig_{N-2}/N}\left(1+A_{N-1}\,P_{N-2}\right)
\label{U}
 \eea
 where $g_k=g_k(r)$ for $k=0,\dots,N-2$ are the profile functions which 
depend only on $r$. Moreover, 
we define also
$A_k=e^{2ig_k}-1$. The boundary value $U \ra I$ at $r \ra \infty$
 (needed for finiteness
of the action) imposes the requirement
that $g_i(\infty)=0$.

 Here $P_k$ form a set of projectors based on the maps $S^2 \ra CP^{N-1}$
 which are constructed as follows \cite{Za}:
write each projector $P$ as
\be
P(V)=\fr{V \otimes V^\dg}{|V|^2},
\label{for}
\ee
where $V$ is a $N$-component complex vector of two variables $z$ and
$\bar{z}$. The first projector is obtained by 
taking $V=f(z)$ (i.e. an analytic
vector of $z$), while the other projectors are given in terms of new vectors $V$
which  are obtained
from the original $V$ by differentiation and Gramm-Schmidt
orthogonalisation.
If we define an operator $P_+$  by its action on any vector $v \in \C^N$
\cite{DinZak} as
\be
P_+ v=\pr_z v- v\,\fr{v^\dg \,\pr_z v}{|v|^2},
\ee
then the vectors $V_k=P^k_+ v$ can be defined by induction:
$P^k_+ v=P_{+}(P^{k-1}_+ v)$.

Therefore, in general, we can consider projectors $P_k$ of the
form (\ref{for}) corresponding to the family of vectors
$V\equiv V_k=P\sp{k}_+f$
(for $f=f(z)$) as
\be
P_k=P(P^k_+ f), \hs \hs k=0,\dots,N-1,
\label{maps}
\ee
where, due to the orthogonality of the projectors, we have 
$\sum_{k=0}^{N-1}P_k=1$. This follows from 
 the following properties of vectors $P^k_+ f$ 
(which hold when $f$ is holomorphic):
\begin{eqnarray}
\label{bbb}
&&(P^k_+ f)^\dg \,P^l_+ f=0 \ \ \ {\rm for} \ \ k\neq l \ , \nonumber\\
&& \pr_{\bar{z}}\left(P^k_+ f\right)=-P^{k-1}_+ f \fr{|P^k_+
f|^2}{|P^{k-1}_+ f|^2} \ , \ \ 
\pr_{z}\left(\fr{P^{k-1}_+ f}{|P^{k-1}_+ f|^2}\right)=\fr{P^k_+
f}{|P^{k-1}_+f|^2} \ .
\end{eqnarray}
Note that, for $SU(N)$,  the last projector in the sequence,  $P_{N-1}$,
corresponds to an anti-analytic vector (i.e. a function of $\bar{z}$).
 Moreover, we can always express one
projector as a sum of the others.

It was shown in \cite{IPZ2} (for the non-gravitating spherical symmetric 
skyrmions) that
 the chiral field (\ref{U}) is an exact  solution of the corresponding equations
when 
\be f=(f_0,...,f_j,...,f_{N-1})^t \ \ \
\mbox{where} \ \ f_j=z^j\sqrt{{N-1}\choose j} \label{smap} 
\ee 
and ${N-1}\choose j$ denote the binomial coefficients.
In what follows we apply the same ansatz to obtain 
 the corresponding gravitating skyrmions  for
the simplest cases of $SU(2)$ and $SU(3)$. 
These cases will clarify the expressions for the general $SU(N)$ case.

\section{Spherical Symmetric equations of motion}

In the case of spherical symmetry,
the action  (\ref{ac})  using (\ref{U}) takes the form
\bea
S&=&2\pi\int\bigg\{\fr{RAr^2}{8\pi G}-\fr{4}{N}\,
r^2CA
\left(\sum_{i=0}^{N-2}g_i'\right)^2+4 r^2 CA
\sum_{i=0}^{N-2}g_i'^2+2A\sum_{k=1}^{N-1}D_k\nonumber\\
&&+\fr{A}{4r^2}\left[D_1^2+\sum_{i=1}^{N-2}\left(D_i-D_{i+1}\right)^2
+D_{N-1}^2\right]+2CA\sum_{k=1}^{N-1}D_k
\left(g_k'-g_{k-1}'\right)^2\bigg\}drdt,
\label{ac2}
\eea
where $D_k=2k(N-k)\,\sin^2 (g_k-g_{k-1})$.
The matter equations are obtained from the variation of this action
with respect to the matter field. We will present these equations
when considering the specific cases of $SU(2)$ and $SU(3)$.

In addition, the Einstein equations (\ref{ff}) take the form:
\bea
\fr{2}{r^2}m'&=&16\pi G\,\bigg[-\fr{C}{N}\left(\sum_{i=0}^{N-2} 
g'_i\right)^2+C\sum_{i=0}^{N-2} g_i^{'2}
+\fr{1}{2r^2}\sum_{k=1}^{N-1}D_k+\fr{C}{2r^2}
\sum_{k=1}^{N-1}D_k\left(g_k'-g'_{k-1}\right)^2
\nonumber\\
&&+\fr{1}{16r^4}\left(D_1^2+\sum_{i=1}^{N-2}\left(D_i-D_{i+1}\right)^2
+D_{N-1}^2\right)\bigg]\acc
\fr{2}{r}\,\fr{A'}{A}C&=&
16\pi G\,\left[-\fr{2C}{N} \left(\sum_{i=0}^{N-2} 
g'_i\right)^2+2C\sum_{i=0}^{N-2} g_i^{'2}
+C\fr{1}{r^2}\sum_{k=1}^{N-1}D_k\left(g'_k-g'_{k-1}\right)^2
\right].
\label{ef}
 \eea 

For simplicity, we set  $F_k=g_k-g_{k+1}$ for $k=0,\dots, N-2$
with $F_{N-2}=g_{N-2}$.
Then, the topological charge (\ref{B}) simplifies to
\be
\label{charge}
B=\fr{1}{\pi}\sum_{i=0}^{N-2}\left(i+1\right)\left(N-i-1\right)
\left(F_i-\fr{\sin 2F_i}{2}\right)
\Bigg|_{0}^{\infty}  \ .
\ee
Since $F_i(\infty)=0$ the only contributions to the topological charge come 
from $F_i(0)$.

The main symmetry of our expressions is the symmetry under the
 independent interchanges
\be
F_k \leftrightarrow F_{N-k-2}, \hs \hs k=0,\dots,N-2
\label{sym}
\ee
which follows from the fact that the $D_k$ terms in (\ref{ac2})
are symmetric under the interchange $D_k \leftrightarrow D_{N-k}$
when $F_{k-1}\leftrightarrow F_{N-k-1}$. At the same time, all the other terms in 
(\ref{ac2})
exhibit this symmetry since they are combinations of $F_i$ and 
their derivatives.

\subsection{$SU(2)$}

For $N=2$ there is only one profile function, $F_0(r)$, and (\ref{ac2}) simplifies to
\be
S=2\pi\int \left\{\fr{RAr^2}{8\pi G}+2r^2CA
F_0'^2+4A\sin^2F_0+2A \fr{\sin^4F_0}{r^2}+4CA
\sin^2F_0 F_0'^2\right\}drdt  \ ,
\ee
while its variation with respect to the profile function gives us 
the matter equation:
\be
\left[CA\,r^2F_0'\left(1+
\fr{2\sin^2 F_0}{r^2}\right)
\right]'
-A\sin 2F_0 \left(1+\fr{\sin^2 F_0}{r^2}
+C\, F_0'^2\right)=0. \label{su2}
\ee
The Einstein equations (\ref{ef}) now take the form:
\bea
\fr{2}{r^2}\,m'&=&8\pi\, G\,
\left[CF_0'^2+
 \fr{2\sin^2 F_0}{r^2}\left(1+C F_0'^2\right)
+\fr{\sin^4 F_0}{r^4}\right] \ , \\
\fr{2}{r}\fr{A'}{A}&=&16\pi\, G\, F_0'^2\left(1+2C\,
\fr{\sin^2 F_0}{r^2}\right).
\label{eu2}
\eea
There is only one solution with the boundary condition
$F_0(0)=\pi$. This solution has topological charge $B=1$ (see (\ref{charge})) and 
has previously been studied in great detail in \cite{DHS,BC}. 
Therefore we will not repeat these calculations here.

\subsection{$SU(3)$}

For $N=3$ there are two profile functions, $F_0(r)$, $F_1(r)$, and
(\ref{ac2}) becomes
\bea
S&=&2\pi\int \bigg\{\fr{RAr^2}{8\pi G}
+\fr{8}{3}\,r^2CA
\left(F_0'^2+F_1'^2+F_0'F_1'\right)
+8A\left(\sin^2F_0+\sin^2F_1\right) \nonumber\\
&+& \fr{8A}{r^2}\left(\sin^4F_0-\sin^2F_0\sin^2F_1+\sin^4F_1\right)
+8CA \left(\sin^2F_0 F_0'^2+\sin^2F_1 F_1'^2\right)
\bigg\}drdt  
\eea
The corresponding equations  for $F_0$ and $F_1$ are now given by:
\begin{equation}
\label{su3_1}
\left[r^2CAF_0'\left(\fr{2}{3}\!+\!
\fr{2\sin^2 F_0}
{ r^2}\right)\!+\fr{r^2}{3}\,CA\,F_1'\right]'\!\!-\!
A\sin 2F_0\left(1+\fr{2\sin^2 F_0-\sin^2 F_1}{r^2}
+CF_0'^2\right)=0 \ ,
\end{equation}
\begin{equation}
\label{su3_2}
\left[r^2CAF_1'\left(\fr{2}{3}\!+\!
\fr{2\sin^2 F_1}
{r^2}\right)\!+\fr{r^2}{3}\,CA\,F_0'\right]'\!\!-\!
A\sin 2F_1\left(1+\fr{2\sin^2 F_1-\sin^2 F_0}{r^2}
+C F_1'^2\right)=0 \ .  
\end{equation}
Note that the above equations are symmetric under the simultaneous
interchange $F_0 \ra F_1$
and $F_1 \ra F_0$.

Finally, the Einstein equations (\ref{ff}) take the form:
\bea
\fr{2}{r^2}\, m'\!&=&\!32\pi \,G\, \bigg[\fr{C}{3}
\left(F_0'^2+F_0'F_1'+F_1'^2\right)+
\fr{1}{r^2}\left(\sin^2 F_0+\sin^2 F_1\right)
+\fr{C}{r^2}\left(\sin^2 F_0 F_0'^2+\sin^2 F_1 F_1'^2\right)
\nonumber\\
&&\hs \hs\hs +\fr{1}{r^4}
\left(\sin^4 F_0- \sin^2 F_0 \sin^2 F_1+\sin^2 F_1\right)\bigg]
\label{su3e} \ , \\
\fr{2}{r}\fr{A'}{A}\!&=&\!64\pi \,G \left[\fr{1}{3}
\left(F_0'^2+F_0'F_1'+F_1'^2\right)+
\fr{\sin^2 F_0}{r^2} F'^2_0+ \fr{\sin^2 F_1}{r^2} F_1'^2\right].
\label{eu3}
\eea
The set of equations (\ref{su3_1})-(\ref{eu3}) can only 
be solved numerically when the right boundary
conditions have been imposed. Similarly to the flat case 
\cite{IPZ2}, we see that there exist
three types of  gravitating  multiskyrmions, which we will discuss in detail
in the following section.

\section{Numerical Simulations}

To solve the equations (\ref{su3_1})-(\ref{eu3}) numerically, 
we have adopted the numerical routine described in 
\cite{acr}.
For convenience, we define  $\alpha^2 = 16 \pi G$ so 
that the flat limit with $C(r)=A(r)=1$ 
corresponds to $\alpha=0$.

\subsection{Boundary conditions}

As will be discussed in more detail in the following,
three kinds of regular finite-energy
solutions exist with the following boundary conditions:
\bea
 \mbox{ (I)}\hs &&   F_0(0) = \pi,\hs \hs \hs F_1(0) = \pi ,\nonumber\\
  \mbox{ (II)}\hs &&   F_0(0) = \pi,\hs \hs \hs F_1(0) = -\pi,\nonumber\\
  \mbox{ (III)}\hs && F_0(0) = \pi,\hs \hs \hs F_1(0) = 0,
\label{no}
\eea
and  for all these cases
\begin{equation}
{\rm (I) \ / \ (II) \ / \ (III)} \ : \ \
F_0(\infty) = 0 \ \ , \ \   F_1(\infty) = 0 \ .
\end{equation}
Furthermore, we have the two supplementary conditions 
for the metric functions:
\begin{equation}
{\rm (I) \ / \ (II) \ / \ (III)} \ : \ \
m(0)=0 \ \ , \ \ A(\infty)=1 \ .
\end{equation}
The condition that $m(r)$ vanishes at the origin $r=0$ is due to 
the requirement of regularity, while the second condition for $A(r)$ results
from the requirement of asymptotic flatness.
The  energy $E$  of the gravitating skyrmions can then be 
determined from  the ``mass function" $m(r)$ at infinity:
\begin{equation}  
E = \frac{4 m(\infty)}{ 3 \pi \alpha^2} \ .
\end{equation} 
With this normalisation, the values of $E$ can be compared to those of the flat limit
\cite{IPZ2}.

Our numerical analysis demonstrates that the three solutions are 
indeed continuously deformed by gravity ($\alpha > 0$).
Next we discuss our numerical results for each set of conditions
given  in (\ref{no}).

\subsection{Case I}

This case corresponds to choosing $F_0(r)=F_1(r)$.
Apart from a trivial rescaling of the coupling constants \cite{KKSO},
the equations for the matter and metric functions are equal
to those of the gravitating $SU(2)$ skyrmion, which were studied in great detail
in \cite{DHS, BC}. However, for completeness, we again present the main
features of these solutions.

The non-gravitating  solution has energy
$E \approx 4.928 = 4 \times 1.232$, i.e. four times the energy
of the $SU(2)$ one-skyrmion. Due to the boundary conditions $F_0(0)=F_1(0)=\pi$
the topological charge is four (see (\ref{charge})). The non-gravitating
solution can thus be interpreted as four noninteracting skyrmions placed
on top of each other in such a way that the baryon (energy) density is
spherically symmetric.

As for the asymptotic behaviour, we notice that the chiral field is of the form
\begin{eqnarray}
    &&  F_0(r)=F_1(r) \approx  \pi - B_{I} r   \ \ \ \ \ {\rm for } \ \ \ \ \ 
0\leq r << 1   \ \ , \nonumber \\
&&      F_0(r)=F_1(r) \approx \frac{\tilde{B}_{I}}{r^2} \ \ \ \ \ {\rm for } \ \ \ \ \
 r >> 1  \ , 
\end{eqnarray}
where $B_{I}$, $\tilde{B}_{I}$ are (shooting) parameters depending on $\alpha$ 
which have to be determined
numerically.

Solving  numerically the system (\ref{su3_1})-(\ref{eu3}) for $\alpha > 0$  
we find that the flat solution is gradually deformed by gravity,
forming a branch of gravitating skyrmions.
In particular, the function $C(r)$ develops a local minimum at 
some intermediate radius~: $r=r_m(\alpha)$, while
the function $A(r)$ has a minimum  $A_{min}=A(0)$
at the origin and then increases 
monotonically.

\begin{figure}
\centering
\epsfysize=18cm
\mbox{\epsffile{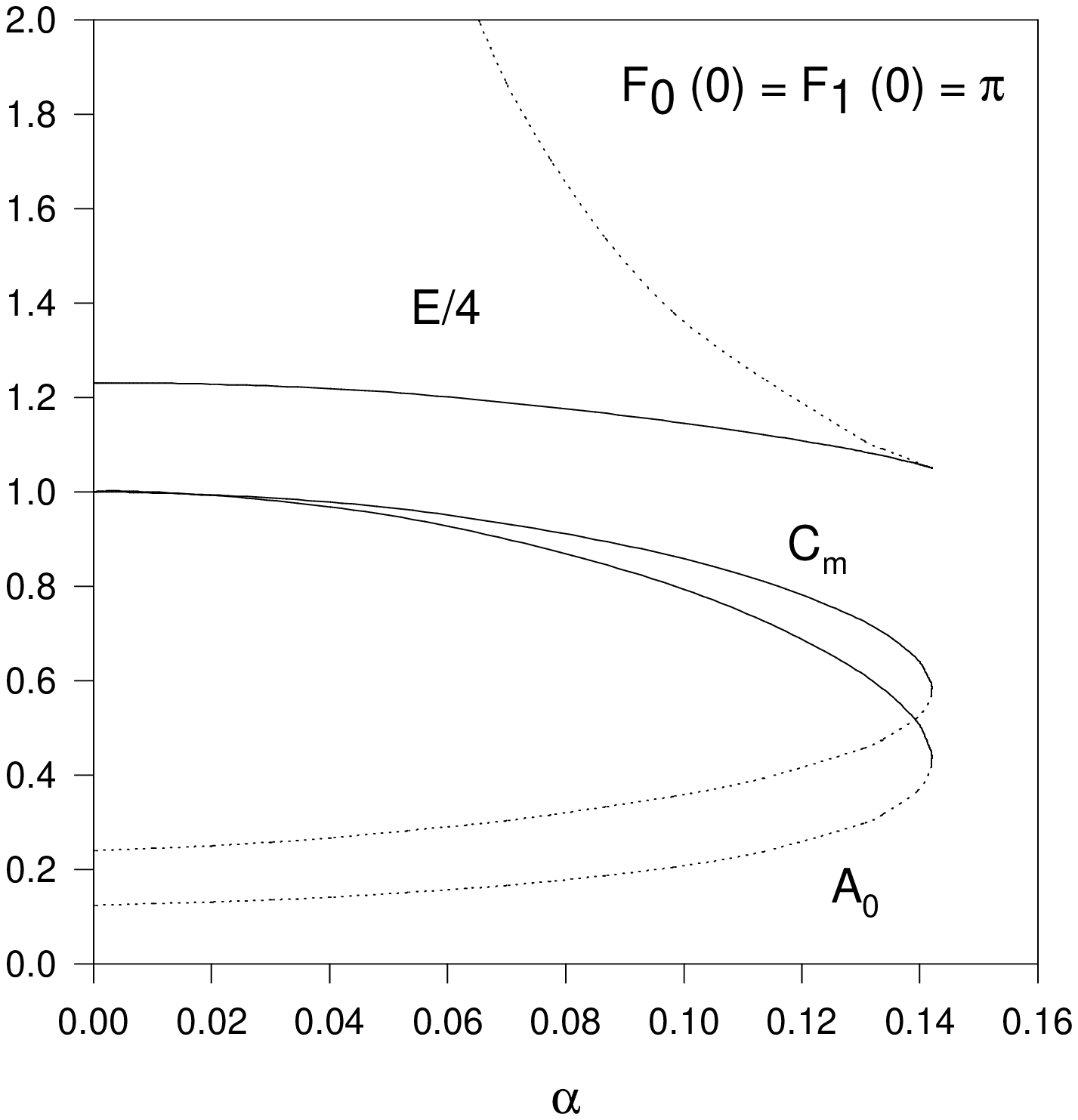}}
\caption{\label{Pr} The values of the metric function $A(r)$ at the origin
$A(0)=A_0$, the minimal value of the metric function $C(r)$, $C_m$ and
the energy $E=\frac{4}{3\pi}\frac{m(\infty)}{\alpha^2}$ of the
gravitating skyrmion solutions for  case I are plotted as functions of $\alpha$.
The solid (respectively dotted) lines refer to the first (respectively  second)
branch of solutions.}
\end{figure}

 However, the profile functions of the Skyrme field
 deviate only slightly from the corresponding ones in the flat limit.
Moreover, as  $\alpha$ increases, the respective minimal values
of the metric functions $C(r)$, $A(r)$, i.e. $C_m =C(r_m)$
and $A_0 = A(0)$ both decrease and so does the corresponding energy $E$.
This latter decrease is, of course, expected since gravity tends to lower
the mass of a solution. The four skyrmions can thus be seen as ``gravitationally
bound''.
This is illustrated in Fig. \ref{Pr} (solid lines).
Note that  this branch of gravitating skyrmions
does not exist for arbitrarily large values
of $\alpha$, but only up to some critical value $\alpha_{cr}$: 
$\alpha \leq \alpha_{cr}\approx 0.142087$
(as shown in Fig.~ \ref{Pr}).
Also the quantities $A_0$, $C_m$ as functions of $\al$
 remain finite with 
\begin{equation}
     E (\alpha=\alpha_{cr}) \approx 4.20,\hs \hs
     A_0 (\alpha=\alpha_{cr})\approx  0.437,\hs \hs
     C_m(\alpha=\alpha_{cr}) \approx  0.584.
\end{equation}

Our numerical analysis, however, strongly suggests that a second branch of solutions
exists in the interval $[0,\alpha_{cr}]$. For a given $\alpha$, the solution
of the second branch has a higher energy than the one on the main branch while
$A_0$, $C_m$ have lower values as shown in Fig. \ref{Pr}
(dotted lines).
For $\alpha \rightarrow \alpha_{cr}$ both branches go to the same
solution but
the solution of the second branch
becomes more and more peaked around the origin (i.e.  the slope $F'(0)$
tends to infinity). 
As a consequence (in this limit)  the energy 
on the upper branch diverges as $\alpha \rightarrow 0$,
but the product $\alpha E$ remains finite ($\approx 0.124$).
This solution, rescaled according to $x \rightarrow \frac{x}{\alpha}$ 
stays regular in the
$\alpha \rightarrow 0$ limit and converges to the sphaleron
solution of the Einstein-Yang-Mills system \cite{KKSO,KUENZLE}.
Note that the metric functions remain finite and are strictly positive.
Therefore,  no black hole solution is generated by the solutions of
the equations  under consideration, in contrast to e.g. the 
Einstein-Yang-Mills-Higgs equations \cite{bfm}.

\subsection{Case II}

By choosing $F_0(r) = - F_1(r)$ with the boundary conditions $F_0(0)=-F_1(0)=\pi$,
the nongravitating solution is topologically trivial since the topological
charge is zero. However, this configuration is not the vacuum solution,
but consists of two skyrmions and two antiskyrmions.
Note that these solutions were not studied previously in \cite{KKSO}.

The regularity of the solutions at the origin requires that $F_0'(0)=0$
which has been checked by us numerically to hold within a very high level of accuracy.
The flat solution ($\alpha=0$) has energy
$E \approx 3.861$ and was discussed in detail in \cite{IPZ2}.

It is worth noticing that the asymptotic 
behaviour  of the Skyrme function $F_0(r)$
differs drastically from the one of case I:

\begin{eqnarray}
&&    F_0(r) \approx  \pi - B_{II} r^2 \ \ \ \ \  {\rm for } \ \ \ \ \ 
0\leq r << 1   \ \ , \nonumber \\
&&      F_0(r) \approx      \frac{\tilde{B}_{II}}{r^3}   \ \ \ \ \ 
{\rm for } \ \ \ \ \ r >> 1 \ .
\end{eqnarray}

For $\alpha >0$ the pattern of the 
solutions is very similar to the one occuring in
case I. The critical value of $\alpha$ is larger than
in I with $\alpha_{cr} \approx 0.1834$. This is related to the
fact that the solutions of case I are heavier (compare the flat limit energies)
and thus exist for a smaller interval of the gravitational coupling.
We note that 
\begin{equation}
     E(\alpha=\alpha_{cr}) \approx 3.377,\hs \hs
     A_0 (\alpha=\alpha_{cr})\approx  0.517,\hs \hs
     C_m (\alpha=\alpha_{cr})\approx  0.614.
\end{equation}

The occurence of two branches is illustrated
in Fig. 2, where the energy $E$ of the solution and 
the quantities $C_m$, $A_0$ are plotted as a function of $\alpha$.

In the limit $\alpha \rightarrow 0$ on the upper branch, the
solution again converge to the SU(3) Einstein-Yang-Mills sphaleron
solution constructed first in \cite{KUENZLE}. 
When solving the equations
numerically, the main feature of the present
charge $0$ solution resides in the fact that the derivative of the chiral
function $F_0(r)$, $F'_0(r)$, vanishes at the origin $r=0$ and
thus the ``shooting'' parameter is  
the second derivative $F_0''(r)|_{r=0}$. This renders the numerical
analysis more involved than in the first case. The solutions' profiles
corresponding to the lower and upper branches are shown in Fig. 3
for $\alpha = 0.17$.
In this figure,
a logarithmic scale is used in order to reveal the completely
different behaviour of $F_0'(r)$ as $r\rightarrow 0$.
The value $F_0''(r)|_{r=0}$ for the solution on the upper branch is several
orders of magnitude larger than its counterpart on the lower branch.
More generally,
we noticed that the shooting parameter $F_0''(r)|_{r=0}$ varies surprisingly
strongly with $\alpha$ and increases considerably when $\alpha$
increases (resp. decreases) on the lower (resp. upper) branch.

\subsection{Case III}
In this case we choose $F_0(0) = \pi$, $F_1(0) = 0$
and $F_1(r)$ goes from zero to zero developping
one node at some finite value  of $r$. Due to the
boundary conditions, these solutions have topological charge $2$ and
are
the gravitating $SO(3)$ embedded solutions discussed in \cite{KKSO}. The
flat limit ($\alpha=0$) solutions have been studied in \cite{Bal}.
Again, for completeness, we discuss the main features of the solutions in
the following.

The regularity of the solutions at the origin  implies 
that $F_0'(0) = F_1'(0)$,
which confirms the validity of  our numerical procedures.
The energy  of the  flat solution is
$E \approx 2.376$. The pattern
with two branches of solutions merging at a critical value of $\alpha_{cr} 
\approx 0.2333$ is also seen here \cite{KKSO}.

The gravitating solution can be characterized by
\begin{equation}
     E (\alpha=\alpha_{cr})\approx 2.033,\hs \hs
     A_0(\alpha=\alpha_{cr}) \approx  0.460 ,\hs \hs
     C_m(\alpha=\alpha_{cr}) \approx  0.537.
\end{equation}

\section{Conclusions}
Our numerical results (for the three cases studied) indicate 
that
the  higher the energy of the solution is, in the flat limit, the smaller is
the interval of $\alpha$ for which the gravitating skyrmions exist. 
The metric functions
of the solutions on the second branch do not 
develop a zero in the critical limit and thus do not represent solutions
with horizons.
It is rather that, in this limit, the matter fields  become singular
at the origin.
However, using a different scale for the radial variable
(e.g. by reinstating the Skyrme coupling constant)
would  render the limiting solutions regular. Following
\cite{km}, we call such solutions ``gravitating sphalerons".

Following the arguments based on Morse theory
(see eg \cite{kus} and references therein), we note that the solutions
on the branch with the higher energy have one  more unstable mode than 
the solutions
on the lower branch. In \cite{KKSO} it was argued that the solutions
on the lower branches in cases I and III are stable (since the flat space
limit on these branches is stable), while the solutions on
the upper branch are unstable. For case II, the flat space solution
on the lower branch is unstable and leads to annihilation \cite{IPZ2}.
Thus, we expect that the gravitating analogues on the lower branch
are unstable and that the solutions on the upper branch have two unstable
modes.

Let us mention that the situation discovered here for the cases I and III
is, in many respects, similar to what
happens with the 
electroweak skyrmion \cite{bhku,eks}.  In the latter case the
starting theory is the standard model of electroweak interactions, in which
the Higgs field plays the role of a Skyrme field and the
relevant coupling constant (call it $\xi$) parametrizes a
supplementary effective interaction (a sort of Skyrme term)
encoding part of the radiative corrections of the theory. 
Two branches
of solutions exist which also merge at a critical value of $\xi$.
The lower branch is stable and is the so-called ``electroweak skyrmion branch''
because it approaches a  skyrmion solution in the limit $\xi \rightarrow 0$.
The higher branch, as shown in  \cite{bhku}, approaches 
either the  sphaleron \cite{km} or (according to the value of the Higgs field
mass) the bisphaleron solution \cite{kubh,ya}.\\

{\bf Acknowledgements} YB acknowledges the Belgian F.N.R.S. for financial
support. TI thanks B Kleihaus for a useful discussion.

\newpage

\newpage
\begin{figure}
\centering
\epsfysize=18cm
\mbox{\epsffile{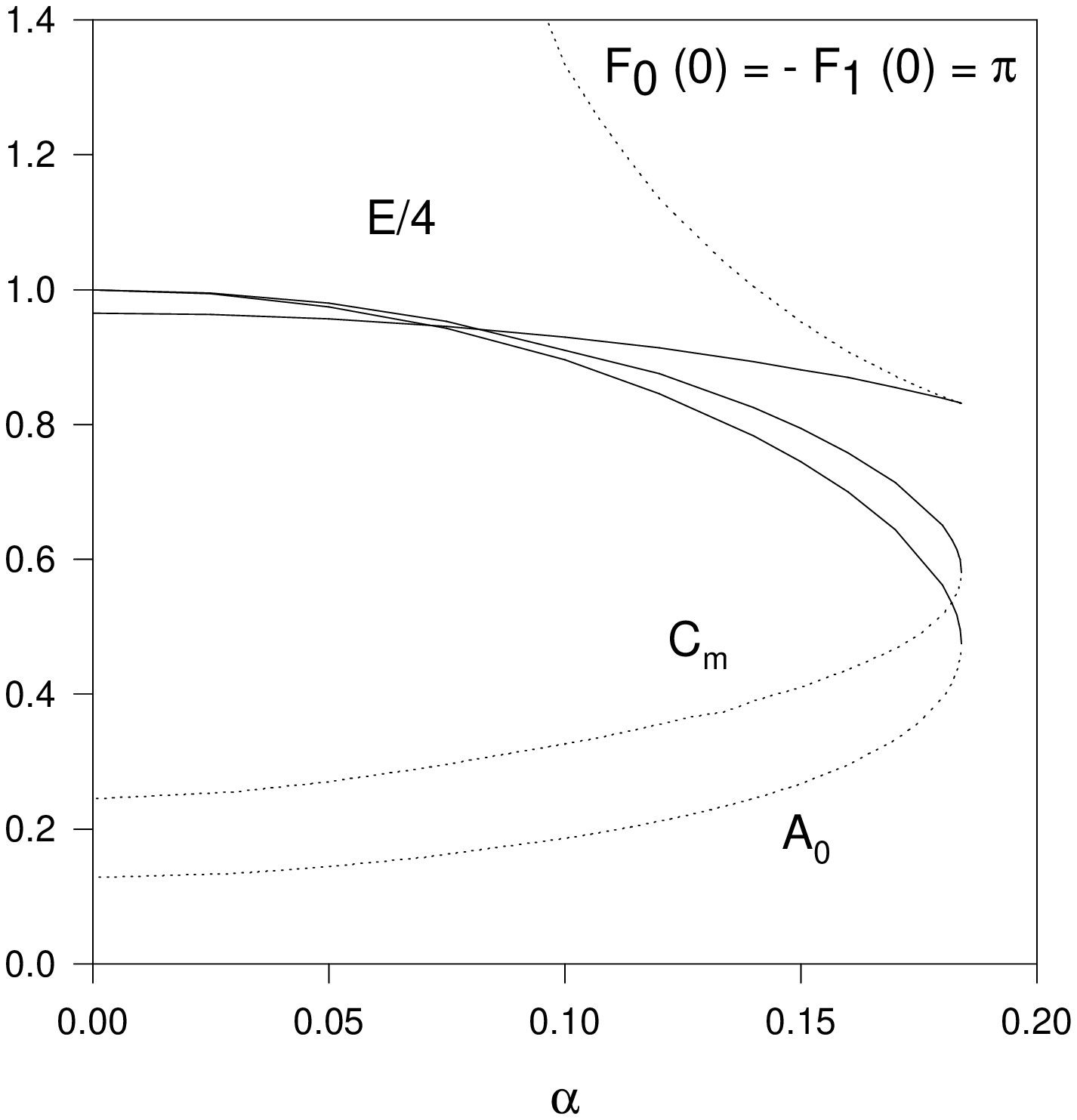}}
\caption{ The values of the metric function $A(r)$ at the origin
$A(0)=A_0$, the minimal value of the metric function $C(r)$, $C_m$ and
the energy $E=\frac{4}{3\pi}\frac{m(\infty)}{\alpha^2}$ of the
gravitating skyrmion solutions for  case II are plotted as functions of $\alpha$.
The solid (respectively dotted) lines refer to the first (respectively  second)
branch of solutions.}
\end{figure}

\newpage
\begin{figure}
\centering
\epsfysize=18cm
\mbox{\epsffile{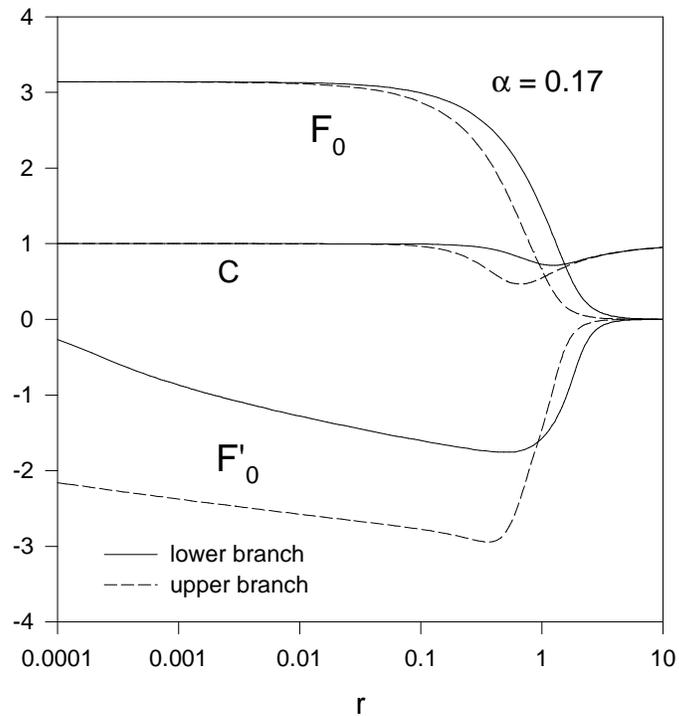}}
\caption{The profiles of the function $F_0(r)$, $F_0'(r)$ and $C(r)$ are shown
for the two branches of skyrmion-antiskyrmion
solutions (case II) with $\alpha=0.17$.}
\end{figure}

\end{document}